\documentstyle[epsf,a4,12pt]{article}

\newcommand{\p}{\partial}
\newcommand{\bra}{\langle}
\newcommand{\ket}{\rangle}

\newcommand{\tr}{\mbox{Tr}\,}
\newcommand{\nn}{\nonumber}

\newcommand{\reseteqnum}{\setcounter{equation}{0}}
\def\dfrac#1#2{\displaystyle\frac{#1}{#2}}

\catcode`\@=11
\newbox\tempboxa
\newdimen\captionboxsubcount
\def\capsize#1{\captionboxsubcount=#1pt}
\newdimen\captionboxsub
\captionboxsub=\hsize \advance\captionboxsub by -\captionboxsubcount
\advance\captionboxsub by -\captionboxsubcount
\long\def\@makecaption#1#2{
 \setbox\@tempboxa\hbox{#1: #2}
 \ifdim \wd\@tempboxa >\captionboxsub
\rightskip=\captionboxsubcount \leftskip=\captionboxsubcount #1: #2
\else \hbox to\hsize{\hfil\box\@tempboxa\hfil}
 \fi}
\catcode`\@=12
\capsize{30}

\begin{document}
%
% ----------------- Title -----------------------
\parindent 0mm
\baselineskip 6mm
\begin{flushright}
KU-AMP 96014 \\
KUNS-1420 \\
HE(TH) 96/15 \\
hep-th/9610244
\end{flushright}
\vspace*{5mm}
\begin{center}
{\Large\bf
Dynamical Gauge Boson \\
\vspace{2mm}
and \\
\vspace{2mm}
Strong-Weak Reciprocity}
\vspace{8mm} \\
{\sc Masako BANDO}
\\
{\it Physics Division \\
Aichi University, Miyoshi, Aichi 470-02, Japan}\footnote{%
E-mail: bando@aichi-u.ac.jp}

{\sc Yusuke TANIGUCHI}
\\
{\it Department of Physics, Kyoto University, Kyoto 606-01, Japan}
\footnote{%
E-mail: tanigchi@gauge.scphys.kyoto-u.ac.jp}
\\
and \\
{\sc Shogo TANIMURA}
\\
{\it Department of Applied Mathematics and Physics \\
Kyoto University, Kyoto 606-01, Japan}\footnote{%
E-mail: tanimura@kuamp.kyoto-u.ac.jp}
\vspace{20mm} \\
\center{\bf Abstract} \\
\end{center}

\baselineskip 5mm
{\small 
\parindent 12pt
It is proposed that asymptotically nonfree gauge theories are
consistently interpreted as theories of composite gauge bosons.
It is argued that when hidden local symmetry is introduced,
masslessness and coupling universality of dynamically generated gauge
boson are ensured.
To illustrate these ideas we take a four dimensional Grassmannian
sigma model as an example and show that the model should be regarded
as a cut-off theory and there is a critical coupling at which the
hidden local symmetry is restored.
Propagator and vertex functions of the gauge field are calculated
explicitly and existence of the massless pole is shown.
The beta function determined from the $ Z $ factor of the 
dynamically generated gauge boson coincides with that 
of an asymptotic nonfree elementary gauge theory.
Using these theoretical machinery we construct a model
in which  asymptotic free and nonfree gauge bosons coexist
and their running couplings are related 
by the reciprocally proportional relation.
}

%----------------------------------------------------------------------
\newpage
\parindent 16pt
\baselineskip 8mm
% \baselineskip 6mm
% ------------------------ section 1 ----------------------
%
\section{Introduction}
Recently it has been demonstrated that two different supersymmetric
gauge theories have
equivalent low energy physics being connected each other by dual
transformation, 
in which a weak coupling system of one theory 
is mapped to a strong coupling system of the other \cite{Seiberg}.
Duality connects two theories which have different gauge groups, 
where, of course, 
the number of gauge bosons and structure of their interaction
are different.
Then there naturally arises a question  
how different theories are transformed each other by duality and what
is dynamical origin of the dual gauge bosons.
That is the question to motivate this investigation.

The idea of dynamically generated gauge bosons has long history
since Bjorken's proposal \cite{Bjorken}, 
in which gauge bosons are to be generated as
bound states of matters, fermions or bosons. 
Originally Bjorken argued that the gauge boson is a
Nambu-Goldstone (NG) boson
responsible for spontaneous breaking of the Lorentz invariance.
Another people \cite{eguchi,suzuki,georgi,hasenfratz} considered a
model with four-fermi vector-vector interaction and used fine-tuning
to impose masslessness of generated vector bosons.
However the most critical assumption of those papers
is that the interaction between matters is very strong
and massless vector bound states appear.
It seems almost impossible to realize the appearance of exact massless
bound states and most papers had to be satisfied 
with an approximate gauge symmetry generation 
where the mass of vector bound states is very small compared 
with the relevant scale (the mass of relevant fermions, for example).

Summarizing points,
although the appearance of vector bound states occurs quite often
in most models, it is difficult to make them massless
and get universal coupling with other matters.
These properties are characteristic for a gauge boson and 
inevitable barrier against dynamical gauge bosons 
if one reminds a critical theorem; It has been proved
\cite{WeinbergWitten, KugoUehara}
that the system must have gauge symmetry
if there is a massless particle with spin $j \ge 1 $.
It is never generated unless gauge symmetry is builtin 
from the first into Lagrangian. 

This has been quite serious 
until the notion of hidden local symmetry came into play
in the context of supergravity theory a l\'a nonlinear sigma model
\cite{cjulia,N=8} (for review see \cite{PhysRep}).
With hidden local symmetry at hand
it is known that in some 2 or 3 dimensional models
gauge fields associated with hidden local symmetry
acquire their own kinetic terms via quantum effects and the poles
of the gauge bosons are developed dynamically \cite{DVL,AA}. 
However these attempts have not been successful in four dimensions 
within renormalizable theories: 
they can be generated only in cut-off theories, 
for example $ CP^{n-1} $ models or Nambu-Jona-Lasinio type models. 
In the following we shall see cut-off theories are enough for our
goal.

Now that hidden gauge symmetry guarantees the appearance of 
dynamical gauge bosons and in the low energy effective theory 
they behave in a similar way as elementary gauge bosons. 
Then
a question arises as to how dynamical gauge bosons 
are discriminated from elementary ones. 
The answer is the compositeness condition and asymptotic nonfree
property.
The idea of compositeness condition at finite scale was first 
argued for the case of dynamical Higgs bosons. 
Bardeen, Hill and Lindner 
\cite{BHL}
proposed that divergence of the yukawa coupling in high energy
indicates that the Higgs bosons are bound states of some elementary
fields.
Compositeness of Higgs bosons is characterized 
by vanishing of its kinetic term, 
namely, vanishing of the wave-function renormalization factor
$ Z = 0 $, which is translated to divergence of the yukawa coupling
$ y = \infty $ by rescaling of the Higgs fields.

Similar argument can be applied to dynamical gauge bosons.
We can interpret divergence 
of running gauge couplings at high energy scale $ \Lambda $ as a
compositeness condition extended for gauge bosons.
Usually in quantum field theories, 
gauge interactions are required to be asymptotically free,
otherwise they become trivial theories
because of the existence of the Landau singularity $ \Lambda_L $. 
We propose that asymptotic nonfree gauge theories are
not trivial but they are to be identified as theories of
dynamical gauge bosons.

In this paper we present a concrete model 
in which dynamical gauge bosons 
are generated as composite vector states whose behavior is 
controlled by the hidden local symmetry. 
In section 2 we investigate a simple example of 
SU($N_f$) / [SU($N_c$) $\times$ SU($N_f-N_c$)] Grassmannian like
model.
We shall show how hidden gauge bosons are generated 
as composite states of NG fields and see how the gauge coupling
runs asymptotically nonfreely, 
until it blows up at $\Lambda$. 
Cooperating the above model with an external gauge field we exhibit
in section 3 another model in which there are 
asymptotically free and nonfree gauge groups.
This model shows a sort of correspondence of 
strong-weak gauge couplings.
Summary and future problems are presented in section~4.

% --------------------------------- section 2 -------------------
%
\section{Compositeness Condition of Gauge Theory}

\subsection{Hidden local symmetry}

Let us explain the notion of hidden local symmetry briefly.
A nonlinear sigma model with $ G/H $-valued fields
has a symmetry $ G_{\rm global} $, which is realized nonlinearly.
These scalar fields describe NG bosons associated with symmetry
breaking from $ G $ to $ H $.
The $ G/H $ nonlinear sigma model is equivalent in classical theory
to a linear sigma model which has a symmetry
$ G_{\rm global} \times H_{\rm local}$.
In the linear model there are $ G $-valued scalar fields
and gauge fields of the gauge group $ H $.
These gauge fields are auxiliary fields
so they do not have their own kinetic terms and can be eliminated.
In this sense one can say that the nonlinear sigma model
has a hidden local symmetry which is brought by introduction of
redundant variables.
Such hidden symmetry is no more than redundancy at classical level.
However once vector bound states are generated at quantum level,
they become independent fields.
Thanks to this hidden gauge symmetry,  
dynamical gauge invariance is guaranteed exactly and 
masslessness and universality of the coupling is ensured
without affected by the theorem \cite{WeinbergWitten,KugoUehara}.

In order to see dynamical generation of gauge bosons we start from the
$ {\rm SU} (N_f) / $ $[ {\rm SU} (N_c) \times {\rm SU} (N_f - N_c) ] $
Grassmannian like model which is equivalent to 
$ [ {\rm SU} (N_f) / {\rm SU} $ $ (N_f-N_c) ] \times 
{\rm SU} (N_c)_{\rm hidden \; local } $ linear model.
This model is constructed with 
$ N_c \times N_f $ complex scalar fields $ \phi_{a i} $ 
and an auxiliary SU$(N_c)$ gauge field $ A_{\mu} $ coupled to the
index $ a = 1 , \cdots , N_c $.
The matter field $ \phi $ is transformed under
$ {\rm SU}(N_f)_{\rm global} \times {\rm SU}(N_c)_{\rm hidden\;local}$
as
$ \phi (x) \to h(x) \phi(x) g^\dagger $
where
$ h(x) \in {\rm SU}(N_c)_{\rm hidden\;local}$ and
$ g \in {\rm SU}(N_f)_{\rm global}$.

The Lagrangian of our model is
\begin{eqnarray}
     {\cal L} &=& 
     \left( D_\mu \phi \right)^\dagger_{i a} 
     \left( D^\mu \phi \right)_{a i}
     - 
     \lambda_{a b} 
     \left( 
         \phi_{b i} \phi^\dagger_{i a} 
         - \frac{N_f}{\omega} \delta_{a b}
     \right)
     + {\cal L}_{\rm GF + FP},
     \label{eqn:lagrangian}
\end{eqnarray}
where
$ \omega $ is a dimensionful coupling
of the nonlinear sigma model.
The covariant derivative $D_\mu \phi$ is written in terms of the
auxiliary gauge field $A_{\mu}$ as
\begin{equation}
D_\mu \phi = \p_\mu \phi - i A_\mu \phi.
\end{equation}
The term ${\cal L}_{\rm GF + FP}$ in (\ref{eqn:lagrangian}) is the
gauge fixing and FP ghost term for $A_\mu$ in which we take the
covariant gauge fixing,
\begin{equation}
{\cal L}_{\rm GF + FP} = -i\delta_{\rm B}\left\{
\tr\left[ \bar{c} \left( \p^\mu A_\mu + \frac{\alpha}{2} B
\right) \right] \right\},
\end{equation}
where $\delta_{\rm B}$ is the BRST transformation.
The hidden local gauge boson $A_\mu$ does not have its kinetic term
and is redundant degrees of freedom of the model.
The $N_c \times N_c$ hermitian scalar field $\lambda_{a b}$ is the
Lagrange multiplier imposing the constraint
\begin{equation}
\phi_{a i} \phi^\dagger_{i b} = \frac{N_f}{ \omega } \delta_{a b}.
\label{eqn:constraint}
\end{equation}
With this constraint $A_\mu$ can be eliminated by substituting the
equation of motion,
\begin{eqnarray}
  A_\mu = -\frac{i\omega}{2N_f}(\p_\mu \phi \, \phi^\dagger 
-  \phi \, \p_\mu \phi^\dagger).
\label{eqn:composite}
\end{eqnarray}
Then we get the following form
\begin{eqnarray}
     {\cal L} &=& 
     \tr \left[ \p_\mu \phi^\dagger
     \p^\mu \phi
     +
     \frac{\omega}{4N_f}
     (\phi^\dagger \p_\mu \phi-\p_\mu \phi^\dagger \,\phi)^2
     \right].
\end{eqnarray}
From this Lagrangian we can reach the original nonlinear sigma model
without hidden local symmetry by fixing the gauge and imposing the
constraint (\ref{eqn:constraint}).

\subsection{Dynamical Gauge Boson}

We can show that the existence of the massless vector mode is
accompanied with the restoration of 
${\rm SU}(N_f)_{\rm global} \times {\rm SU}(N_c)_{\rm hidden\;local}$
symmetry as in the abelian case if we assume the
deconfining phase \cite{Kugo-Townsend,KTU}.\footnote{
This assumption is justified as we shall show later that
the generated gauge boson interacts weakly in the low
energy region.
}
This can be easily seen by the argument of
the identity;
\begin{equation}
{\rm F.T.}i\bra D_\mu c^A(X) \bar{c}^B(y) \ket
= -\delta^{A B} \frac{p_\mu}{p^2},
\label{eqn:WT}
\end{equation}
and the BRST transformation of $A_\mu$;
\begin{equation}
\left[ iQ_{\rm B} , A_\mu^A \right] = D_\mu c^A.
\label{eqn:BRST}
\end{equation}
Equation (\ref{eqn:WT}) implies that $D_\mu c$ contains the
massless asymptotic field, and so does $A_\mu$ with the help of
equation (\ref{eqn:BRST}).
There are two alternatives to identify the massless mode in $A_\mu$ as
\begin{enumerate}
\item the Nambu-Goldstone mode of broken SU($N_c$) local symmetry,
\item the longitudinal mode of unbroken SU($N_c$) gauge boson.
\end{enumerate}
In the broken phase (1), the breaking of SU($N_c$) local
symmetry by the V.E.V. of
$\bra \phi \ket = \sqrt{N_f} v \left ( \delta_{a b} \>,\> 0\right)$
is accompanied with the breaking of SU$(N_f)$ symmetry.
On the other hand, in the symmetric phase (2), $A_\mu$
is a massless vector boson and any symmetry should not be broken.

The phase of the model is determined by the effective potential.
Let us take the V.E.V.s of $\phi$ and
$ \lambda $ as follows to leave SU$ (N_c) \times $SU($N_f-N_c$)
symmetry,
\begin{eqnarray}
\bra \phi \ket &=& \sqrt{N_f} \, v \, (\delta_{a b} , 0 ),\\
\bra \lambda \ket &=& \lambda \, \delta_{a b}.
\end{eqnarray}
By substituting them the effective potential is given in the $1/N_f$
approximation as
\begin{equation}
        V_{\rm eff} 
        = N_f N_c 
        \left[
                \lambda 
                \left( v^2 - \frac{1}{ \omega } \right)
                + \int \frac{d^4 k}{i(2\pi)^2} \ln (k^2-\lambda) 
        \right].
\end{equation}
This potential coincides with that of $CP^{N-1}$ model
\cite{PhysRep,KTU} within this approximation
and the ground state is determined by the equations
\begin{eqnarray}
        &&
        \frac{1}{N_f N_c} \frac{\p V_{\rm eff}}{\p \lambda}
        = 
        v^2 - \frac{1}{ \omega } 
        + \int \frac{d^4 k}{i(2\pi)^4} \frac{1}{\lambda-k^2} 
        = 0,
        \label{stationary}
        \\
        &&
        \frac{1}{N_f N_c} \frac{\p V_{\rm eff}}{\p v}
        = 
        2\lambda v = 0.
\end{eqnarray}
Here we introduce a cut-off $ \Lambda $
to define the integration in (\ref{stationary}), which is rewritten as
\begin{eqnarray}
  v^2 - f ( \Lambda, \, \lambda)
  = \frac{1}{ \omega_{\rm r} ( \mu ) } 
  - \frac{1}{ \omega_{\rm cr}(\Lambda , \, \mu )}
\label{cutoff}
\end{eqnarray}
where
\begin{eqnarray}
&&
\frac{1}{\omega_{\rm r}}
\equiv
\frac{1}{ \omega } 
- \int \frac{d^4 k}{i(2\pi)^4} \frac{1}{\mu^2 - k^2}
\\
&&
\frac{1}{\omega_{\rm cr}}
\equiv
\int \frac{d^4 k}{i(2\pi)^4} \frac{1}{      - k^2}
- \int \frac{d^4 k}{i(2\pi)^4} \frac{1}{\mu^2 - k^2}
\\
&&
f(\Lambda , \, \lambda)
\equiv
\int \frac{d^4 k}{i(2\pi)^4} \frac{1}{        - k^2}
- \int \frac{d^4 k}{i(2\pi)^4} \frac{1}{\lambda - k^2}.
\end{eqnarray}
$ \omega_{\rm r} $ is a redefined coupling
at the renormalization point $ \mu $.
If the dimension is less than four
$\omega_{\rm r}$, $ \omega_{\rm cr} $ and $ f $ remain finite
in the limit of $ \Lambda \to \infty $,
namely, the theory is renormalizable
as known before~\cite{PhysRep,KTU}.
However, in four dimensions they become divergent,
so Eq. (\ref{cutoff}) makes sense
only for the finite cut-off $ \Lambda $.
The function $ f $ is understood 
as a non-negative function of $ \lambda $
vanishing at $ \lambda = 0 $.
The critical coupling $ \omega_{\rm cr} $ separates two phases
\begin{enumerate}
\item broken phase : 
        $ \lambda = 0, \;\; v \neq 0$, 
        \quad when $ \omega_{\rm r} < \omega_{\rm cr} $,
\item symmetric phase : 
        $ \lambda \neq 0, \;\; v = 0 $,
        \quad when $ \omega_{\rm r} > \omega_{\rm cr} $.
\end{enumerate}
The critical coupling $ \omega_{\rm cr} $
is reciprocally proportional to the logarithmic divergence 
$ \omega_{\rm cr}^{-1} \sim \log \Lambda $,
so it becomes smaller for the larger cut-off $ \Lambda $.
Therefore the symmetric phase is always realized by taking enough
large $ \Lambda $.
In this phase the scalar field $\phi$ is massive
($m^2=\lambda$) whereas the vector $A_\mu$ becomes massless.
Hence this composite vector field (\ref{eqn:composite}) is stable.
In the broken phase $\phi$ field becomes massless
(NG bosons associated with the broken symmetries), 
while the gauge field $A_\mu$, if it exists, becomes massive by the
Higgs mechanism.
This massive gauge boson is unstable as it can decay into two
$\phi$ bosons.

\begin{figure}[htbp]
\epsfxsize=9cm
\begin{center}
\ \epsfbox{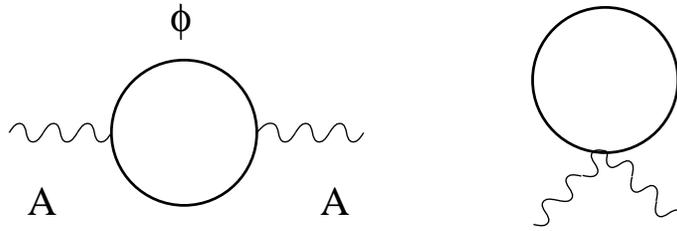}
\vspace{-5pt}
\caption[]{
The Feynman diagrams 
which generate the propagator of the dynamical gauge boson.
}
\label{fig:Z0}
\end{center}
\end{figure}

\begin{figure}[htbp]
\epsfxsize=11cm
\begin{center}
\ \epsfbox{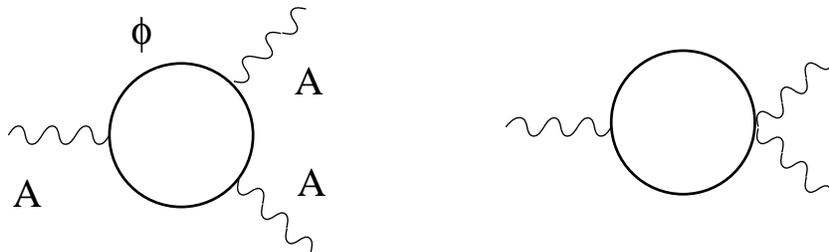}
\vspace{-5pt}
\caption[]{
The Feynman diagrams which generate the three-point self-interaction.
}
\label{fig:Z1}
\end{center}
\end{figure}

\begin{figure}[htbp]
\epsfxsize=14cm
\begin{center}
\ \epsfbox{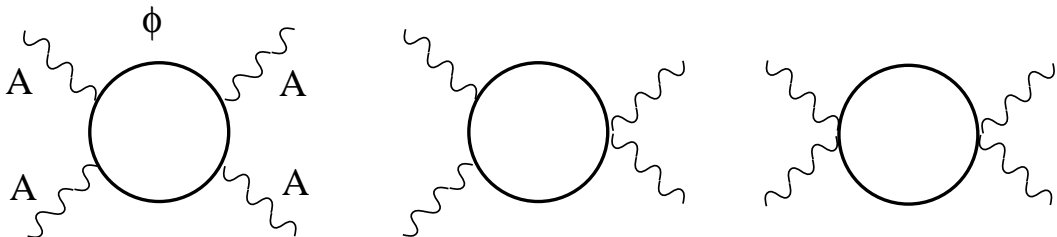}
\vspace{-5pt}
\caption[]{
The Feynman diagrams which generate the four-point self-interaction.
}
\label{fig:Z2}
\end{center}
\end{figure}

To see the generation of massless gauge boson explicitly in
the symmetric phase we calculate the effective Lagrangian of the
composite gauge boson.
We calculate the Feynman diagrams
fig.\ref{fig:Z0}, fig.\ref{fig:Z1}, fig.\ref{fig:Z2}
which contribute to the leading order terms of expansion with respect
to $N_c/N_f$.
The vertex functions are given by
\begin{eqnarray}
        &&
        \Gamma^{(2)} (p) 
        = -\frac{1}{16\pi^2} \frac{1}{6} N_f 
        \ln (\frac{\Lambda^2}{\mu^2})
        \, \delta^{AB} (g_{\mu \nu} p^2 - p_\mu p_\nu) ,
        \label{2point}
        \\
        &&
        \Gamma^{(3)} (k,p,q) 
        = -\frac{1}{16\pi^2}
        \frac{1}{6}  N_f \ln (\frac{\Lambda^2}{\mu^2})
        \, i f^{ABC} 
        [ g_{\mu \nu} (k-p)_\rho 
        \nn\\
        &&  
        \qquad \qquad \qquad \qquad 
        \qquad \qquad \qquad \qquad \quad
        + g_{\nu \rho} (p-q)_\mu
        + g_{\mu \rho} (q-k)_\nu ],
        \\
        &&
        \Gamma^{(4)} (k,l,p,q) 
        = -\frac{1}{16\pi^2} 
        \frac{1}{6} N_f \ln (\frac{\Lambda^2}{\mu^2})
        [
          f^{ABE}f^{ECD} (g_{\alpha \gamma} g_{\beta \delta}
        - g_{\alpha \delta} g_{\beta \gamma})
        \nn\\
        && 
        \qquad \qquad \qquad \qquad 
        \qquad \qquad \qquad \quad 
        + f^{ACE}f^{EBD} 
        ( g_{\alpha \beta}  g_{\gamma \delta}
        - g_{\alpha \delta} g_{\beta \gamma})
        \nn\\
        &&
        \qquad \qquad \qquad \qquad 
        \qquad \qquad \qquad \quad 
        + f^{ADE}f^{EBC} 
        ( g_{\alpha \beta}  g_{\gamma \delta}
        - g_{\alpha \gamma} g_{\beta \delta})
        ].
\end{eqnarray}
And they are put into the effective Lagrangian
\begin{eqnarray}
{\cal L} &=&
-\frac{1}{4} Z_0 \left(\p_\mu A_\nu^A - \p_\nu A_\mu^A \right)^2\nn\\
&& -\frac{1}{2} Z_1 f^{ABC} (\p_\mu A_\nu^A - \p_\nu A_\mu^A)
A_\mu^B A_\nu^C
-\frac{1}{4} Z_2 f^{EAB} f^{ECD} A_\mu^A A_\nu^B A_\mu^C
A_\nu^D\nn\\
&& + \tr \left( |\p_\mu \phi -i A_{\mu} \phi|^2
- \lambda |\phi|^2 \right),
\label{eqn:effective-lagrangian}
\end{eqnarray}
with $Z$ factors;
\begin{eqnarray}
Z_0 = Z_1 = Z_2 = \frac{1}{16\pi^2} 
\frac{1}{6} N_f \ln (\frac{\Lambda^2}{\mu^2}).
\end{eqnarray}
The $Z$ factors in (\ref{eqn:effective-lagrangian}) are dependent on
$\Lambda$.
And we redefine the field $A_{{\rm r}\,\mu}$ so as to normalize the
gauge kinetic term in (\ref{eqn:effective-lagrangian}),

\begin{eqnarray}
A_{{\rm r}\,\mu} &=& \sqrt{Z_0} A_\mu.
\end{eqnarray}
Then the renormalized coupling is determined as
\begin{eqnarray}
g_{\rm r} &=& \frac{1}{\sqrt{Z_0}} = \frac{Z_1}{(\sqrt{Z_0})^3}
= \frac{\sqrt{Z_2}}{Z_0},
\label{eqn:renormalized_coupling}
\end{eqnarray}
which confirms universality of the coupling.
It must be noted that the $Z$ factors vanish at $\mu = \Lambda$.
This means the dynamically generated kinetic and interaction terms in
(\ref{eqn:effective-lagrangian}) disappear,
which we call compositeness condition at the cut-off $\Lambda$.
In conventional normalization 
we always take $ Z(\mu)=1 $ at any scale $ \mu $
to normalize the kinetic term.
This, in turn, indicates that the running coupling $g_r$ 
in (\ref{eqn:renormalized_coupling})
becomes infinity at scale $\Lambda$, 
implying that the theory is asymptotically nonfree gauge theory.
The beta function obtained from (\ref{eqn:renormalized_coupling})
properly reflects this asymptotic nonfree behavior;
\begin{eqnarray}
\beta(g_{\rm r}) = \mu \frac{\p}{\p\mu} g_{\rm r}
= \frac{g_{\rm r}^3}{16\pi^2} \frac{1}{6}N_f.
\label{eqn:beta-fn}
\end{eqnarray}

\subsection{Compositeness Condition}

It is easy to confirm that the beta function (\ref{eqn:beta-fn}) 
of the composite theory coincides with that of the elementary theory
(the low energy theory with elementary gauge field);
\begin{eqnarray}
        &&
        {\cal L} 
        = 
        \tr \left(
          |D_\mu \phi|^2 - \lambda | \phi |^2
        \right)
        - \frac{1}{4}\tr (F_{\mu \nu})^2
        + {\cal L}_{\rm FP+GF},
        \label{eqn:elementary-lagrangian}
        \\
        &&
        \beta(g_{\rm r}) = \frac{g_{\rm r}^3}{16\pi^2} 
        \left( \frac{1}{6}N_f - \frac{11}{3}N_c \right),
\label{eqn:coupling-el}
\end{eqnarray}
in $N_c/N_f$ expansion.
With this approximation the first term in (\ref{eqn:coupling-el})
dominates (the beta function is positive) and shows asymptotically
nonfree character.
As is well known asymptotically nonfree gauge theory necessarily has
Landau singularity at the point $\Lambda_L$ where the running coupling
blows up and is thought to be nonsense as a field theory.
But as is discussed in the previous section this singularity can be
interpreted as $Z=0$ at $\Lambda_L$ in the composite theory.
This $\Lambda_L$ is nothing but the cut-off of nonlinear sigma
model where the kinetic as well as self coupling terms of $A_\mu$
disappear 
and the gauge field loses its identity as an elementary particle.
The asymptotically nonfree scalar gauge theory
(\ref{eqn:elementary-lagrangian}) can be regarded as dynamically
generated gauge theory.

It is indeed always true that the positive beta function of the
composite theory appears in the leading order of $ N_c / N_f $.
However once one proceeds to the next order,  
contribution from non-abelian interactions (the second term in
(\ref{eqn:coupling-el})) can not be neglected.
Although calculation of next order terms is quite difficult,
we may estimate them from the elementary theory using usual RGE
technique.
This situation is similar to the cases of dynamical
Higgs bosons where we got the information of next order
contributions coming from Higgs loops
by looking at the coefficients of beta functions
of elementary theory \cite{composite,composite2}. 

It should be noticed
if the beta function becomes negative due to the next order terms
the theory does not satisfy the compositeness condition but is to be
understood as an elementary gauge theory without cut-off.
 For this case the $N_c/N_f$ expansion is no more applicable.

%
% -------------------------- section 3 -------------------
\reseteqnum

\section{Strong-Weak Reciprocity}

\subsection{Model}

In the previous section we have established a consistent
mechanism to generate a dynamical massless gauge boson.
Its compositeness is characterized by 
the asymptotically nonfree running coupling.
In this section we would like to use this mechanism to construct
a model which exhibits a feature of strong-weak duality.

Namely, the model has two kinds of gauge fields;
one has an asymptotically free coupling
and the other has an asymptotically nonfree coupling.
In low energy region the asymptotically nonfree gauge boson
has a weak coupling and behaves almost freely, 
so a perturbation theory in terms of this boson describes 
dynamics of the model well,
while the asymptotically free gauge boson has a diverging coupling, 
then its perturbative description becomes inadequate.
On the other hand, in high energy region
the asymptotically nonfree gauge boson has a diverging coupling,
reflecting its compositeness.
Then the perturbative method does not work well.
In turn, the asymptotically free gauge boson behaves as a perturbative
constituent.
In this sense, the model provides an example in which 
two gauge fields coexist and give complementary
description of dynamics; one works in low energy
and the other works in high energy.
Such a model is to be constructed below.

Constituents of the model are almost same 
to those of the previous model.
$ \phi = ( \sigma, \pi ) $ are $ N_c \times N_f $ 
complex scalar fields;
$ \sigma = ( \sigma_{ai} ) $ are $ N_c \times N_c $ complex scalars;
$ \pi = ( \pi_{a \alpha} ) $ are $ N_c \times ( N_f - N_c ) $
complex scalars;
the matrix formed by $ ( \sigma, \pi ) $ is denoted as $ \phi $.
$ A_\mu $ is an SU$ (N_c) $ gauge field coupled to the index 
$ a = 1, \cdots, N_c $;
$ V_\mu $ is an SU$ ( N_f - N_c ) $ gauge field coupled
to the index $ \alpha = 1, \cdots, N_f - N_c $;
$ D_\mu \phi 
= \partial_\mu \phi - i g_1 A_\mu \phi + i g_2 \phi V_\mu $.
With these constituents the model is defined by
\begin{eqnarray}
        {\cal L}
        & = &
        \tr
        \left(
                D_\mu \phi^\dagger D^\mu \phi 
                - \lambda ( \phi \, \phi^\dagger - N_f v^2 )
        \right)
        - \frac14 \tr ( F^A_{\mu \nu} )^{2} 
        - \frac14 \tr ( F^V_{\mu \nu} )^{2} 
        \nonumber \\
        &&
        - i \delta_{1 {\rm B}} \tr
        ( \bar{c}_1
        ( \partial^\mu A_\mu + \frac{\alpha_1}{2} B_1 )
        )
        - i \delta_{2 {\rm B}} \tr
        ( \bar{c}_2
        ( \partial^\mu V_\mu + \frac{\alpha_2}{2} B_2 )
        ),
\end{eqnarray}
where $ \delta_{i {\rm B}} \: ( i=1, 2 ) $ is the BRST transformation 
for each gauge group.
This model is similar to (\ref{eqn:elementary-lagrangian}) except that
another gauge field $ V_\mu $ and its kinetic term are introduced
from the beginning.
%
% ----------------------------------------------------------
%
\subsection{Renormalization Group Equations}
The one-loop beta functions of this model are given by
\begin{eqnarray}
        &&
        \beta_1 ( g_1 )
        = \mu \frac{ \partial g_1 }{ \partial \mu }
        = \frac{ g_1^3 }{ 16 \pi^2 }
                \left(
                        \frac{1}{6} N_f - \frac{11}{3} N_c
                \right),
        \label{model2;beta1}
        \\
        &&
        \beta_2 ( g_2 )
        = \mu \frac{ \partial g_2 }{ \partial \mu }
        = \frac{ g_2^3 }{ 16 \pi^2 }
                \left(
                        \frac{1}{6} N_c
                        - \frac{11}{3}( N_f - N_c )
                \right)
        \nonumber \\
        &&
        \qquad \qquad 
        =  \frac{ g_2^3 }{ 16 \pi^2 }
                \left(
                       \frac{23}{6} N_c - \frac{11}{3} N_f 
                \right).
        \label{model2;beta2}
\end{eqnarray}
The particularly interesting situation is that
$ g_1 $ is asymptotically nonfree and
$ g_2 $ is asymptotically free,
namely, the region $ N_c < \dfrac{1}{22}N_f $.
In this region the gauge boson $ A_\mu $ satisfies the compositeness
condition and can be interpreted as a dynamical gauge boson,
which is generated from the Grassmannian like model as shown
in the previous section.
While $ V_\mu $ is an elementary gauge boson,
which is newly introduced here by gauging a subgroup SU $(N_f - N_c )$
of the global SU$(N_f)$.
The $ N_c / N_f $ expansion leads
to the same effective action of $ A_\mu $ as in the previous section,
as far as
the newly introduced coupling $ g_2 $ is of order $ N_c / N_f $.

Now we examine behavior of running couplings closely.
A typical one-loop beta function
\begin{equation}
        \beta(g) 
        = \frac{ \partial g }{ \partial \ln \mu }
        = \frac{b}{ 16 \pi^2 } g^3
\end{equation}
is integrated to give
\begin{equation}
        \frac{1}{g^2(\mu)} - \frac{1}{g^2(\Lambda)}
        =
        \frac{b}{ 8 \pi^2 } \ln ( \Lambda / \mu ).
\end{equation}
Putting $ \alpha = \dfrac{g^2}{ 4 \pi } $ it is rewritten as
\begin{equation}
        \alpha( \mu )
        =
        \frac{1}{ \dfrac{1}{\alpha( \Lambda )}
        + \dfrac{b}{2 \pi} \ln( \Lambda / \mu ) }
        =
        \frac{2 \pi}{b \ln( \Lambda / \mu )},
\end{equation}
where we take $ \alpha( \Lambda ) = \infty $.
If $ b > 0 $, the gauge coupling is well-defined only
in the lower energy region $ \mu < \Lambda $
and $ g $ diverges as $ \mu $ approaches to $ \Lambda $.

\begin{figure}[htbp]
\epsfxsize=10cm
\begin{center}
\ \epsfbox{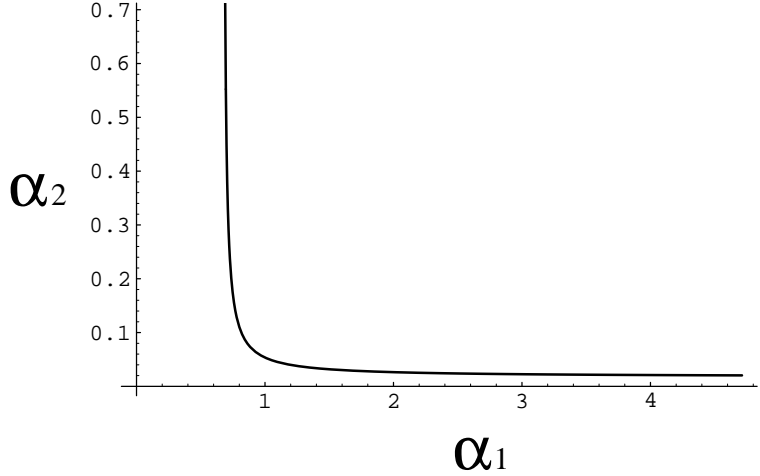}
\vspace{-5pt}
\caption[]{
The reciprocal relation between $\alpha_1$ and $\alpha_2$
for $N_f=100$ and $N_c=2$.
}
\label{fig:reciprocity}
\end{center}
\end{figure}

 For $ \alpha_1 = \dfrac{g_1^2}{ 4 \pi } $ and
$ \alpha_2 = \dfrac{ g_2^2 }{ 4 \pi } $ we have
\begin{equation}
        \frac{ 1 }{ b_i \alpha_i (\mu) } 
        = \frac{1}{ 2 \pi } \, \ln ( \Lambda_i /  \mu ).
        \qquad ( i = 1, \, 2 ) 
\end{equation}
Elimination of $ \ln \mu $ gives a renormalization-invariant relation
between $ \alpha_1 $ and $ \alpha_2 $, that is
\begin{equation}
        \frac{ 1 }{ b_1 \alpha_1 (\mu) } 
        - 
        \frac{ 1 }{ b_2 \alpha_2 (\mu) } 
        =
        \frac{1}{ 2 \pi } \, \ln \frac{ \Lambda_1 }{ \Lambda_2 }.
        \label{reciprocity}
\end{equation}
Typical feature of this equation is shown in the figure
\ref{fig:reciprocity}.

\begin{figure}[htbp]
\epsfxsize=10cm
\begin{center}
\ \epsfbox{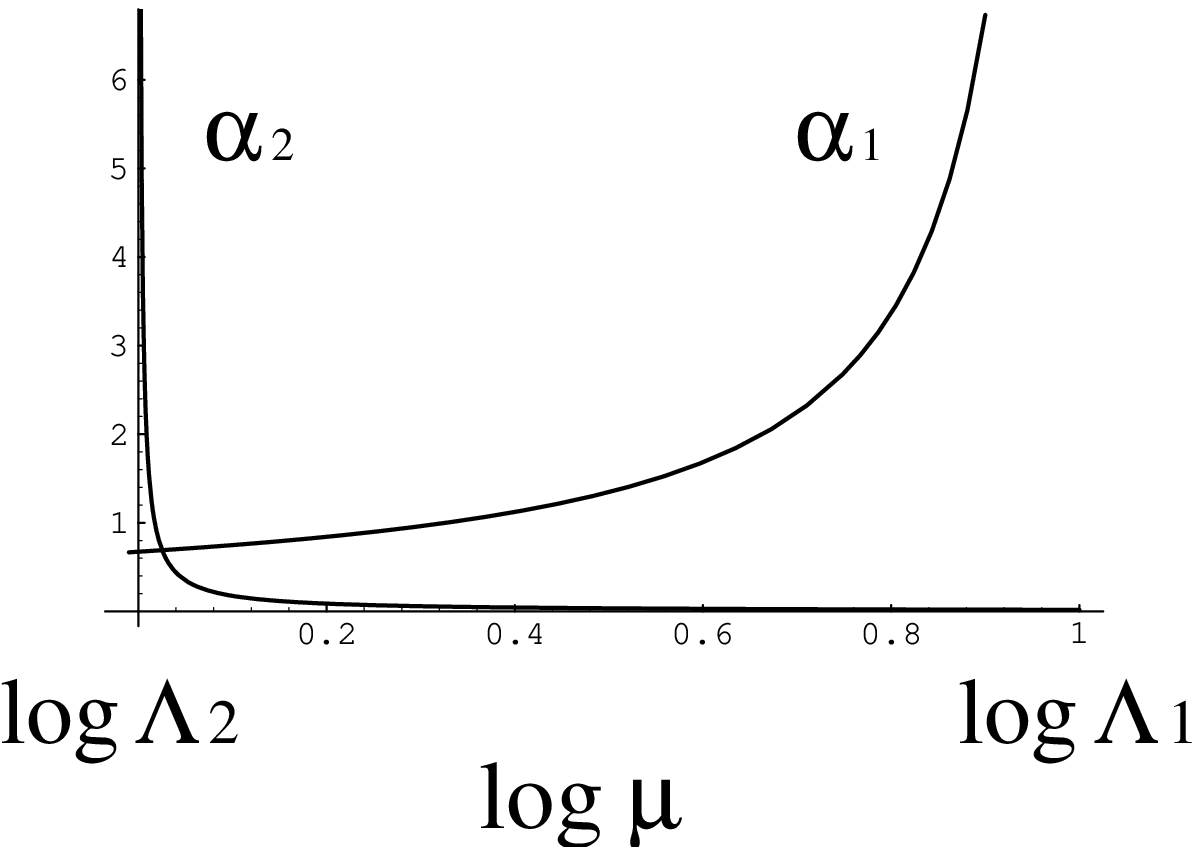}
\vspace{-5pt}
\caption[]{
Running couplings for $ N_f=100 $ and $ N_c=2 $ (region i).
The energy is scaled in the unit of $\Lambda_2$.
}
\label{fig:region1}
\end{center}
\end{figure}

\begin{figure}[htbp]
\epsfxsize=10cm
\begin{center}
\ \epsfbox{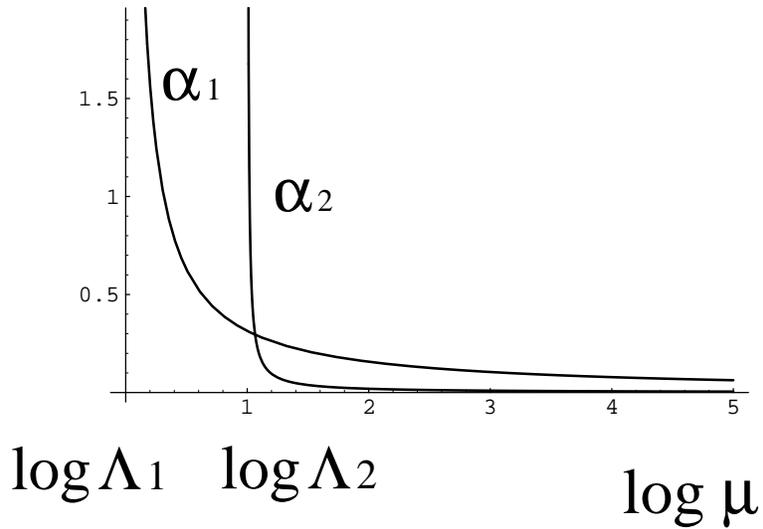}
\vspace{-5pt}
\caption[]{
Running couplings for $ N_f=100 $ and $ N_c=10 $ (region ii).
The energy is scaled in the unit of $\Lambda_1$.
}
\label{fig:region2}
\end{center}
\end{figure}

There are three cases:
\begin{enumerate}
\renewcommand{\labelenumi}{(\roman{enumi})}%
\item   $ N_c < \dfrac{1}{22} N_f $:
        At this time $ b_1 > 0 $, $ b_2 < 0 $, 
        then $ g_1 $ runs asymptotically nonfreely
        and $ g_2 $ runs asymptotically freely
        in the region $ \Lambda_2 < \mu < \Lambda_1 $
        (fig.\ref{fig:region1}).
\item $ \dfrac{1}{22} N_f < N_c < \dfrac{22}{23} N_f $:
        $ b_1, \, b_2 < 0 $,
        then both $ g_1 $ and $ g_2 $ run asymptotically freely
        in the region $ \mu > \Lambda_1, \,  \Lambda_2 $ 
        (fig.\ref{fig:region2}).
\item   $ \dfrac{22}{23} N_f < N_c $:
        $ b_1 < 0 $, $ b_2 > 0 $,
        then $ g_1 $ runs asymptotically freely
        and $ g_2 $ runs asymptotically nonfreely
        in the region $ \Lambda_1 < \mu < \Lambda_2 $.
        The situation is the reverse of the first case.
\end{enumerate}
The first case is the most interesting, which shall be discussed 
in detail below.
SU$ ( N_c ) $ and SU$ ( N_f - N_c ) $ gauge bosons coexist
at the scale $ \mu $ in the region $ \Lambda_2 < \mu < \Lambda_1 $.
There the compositeness condition for SU$ (N_c) $ gauge 
boson is satisfied.
Notice that the beta function (\ref{model2;beta1}) coincides with
Eq. (\ref{eqn:beta-fn}) if $ N_c \ll N_f $.

As the scale $ \mu $ approaches to $ \Lambda_1 $,
the coupling $ g_1 $ diverges
and the other coupling $ g_2 $ becomes small.
It implies that
the compositeness of the SU$ ( N_c ) $ gauge boson becomes apparent
and it does not behave as an elementary particle.
The smallness of $ g_2 $ ensures that perturbation
of SU$ ( N_f - N_c ) $ gauge boson
provides a good description of the model in high energy.

On the contrary,
as the scale comes down to $ \Lambda_2 $,
the situation is reversed.
Now $ g_2 $ runs to infinity, 
that is so-called infrared slavery.
The SU$ ( N_f - N_c ) $ gauge boson is strongly interacting
and does not provide a good perturbative description.
However
SU$ ( N_c ) $ gauge boson is rather weakly interacting
and then works as a perturbative constituent.
The observed complementary role of two gauge bosons and
the reciprocal relation between two couplings (\ref{reciprocity})
may suggest a mechanism to understand duality.

%
% -------------------------------- section 4 ----------------------
%
\section{Discussion}

The idea of dynamical generation of particles is not new; 
actually Bardeen-Hill-Lindner \cite{BHL} 
had noticed dynamical Higgs whose yukawa coupling blows up 
at some high energy scale $\Lambda$. 
However dynamical gauge bosons have long been thought to be 
quite different because of the existence of gauge symmetry.

By investigating the Grassmannian like model in
which non-abelian gauge group is included, 
we have observed 
%that dynamical gauge boson behaves asymptotically nonfreely.
that the dynamical gauge boson appears to be asymptotically nonfree.
Masslessness and coupling universality
of dynamically generated gauge boson are ensured 
by virtue of the hidden local symmetry.

This indicates that asymptotically nonfree gauge theories,
which have been thought to be nonsense as a field theory,
are consistently interpreted as theories of composite gauge bosons. 
The beta function determined from the $ Z $ factor of the 
dynamically generated gauge boson turns out to coincide with that
of the elementary but asymptotic nonfree gauge boson.

Using these theoretical machinery we constructed a model
which simulates duality of supersymmetric gauge theories.
In this model asymptotically free and nonfree gauge bosons coexist
and their running couplings are related 
by the reciprocally proportional relation.
One of the motivations of this work has been to find a mechanism
which relates dual gauge bosons.
While in the model which we have constructed here,  
dual gauge bosons are not transformed each other
but they coexist.
In this sense their relation is not literally duality. 
However, so far as we concern effective theory in extremely low
or high energy regions near $ \Lambda $'s, the system is governed by
either of the coexisting gauge bosons.
In this sense they play dual roles according to the relevant energy
scale and the number of matter contents. 

So far as we are concerned with the low energy effective theory, 
it is enough to deal it within the framework of cut-off theory.
In this paper we find the connection between the low energy effective
theory below $\Lambda$ with the theory near $\Lambda$.
Above this cut-off we may expect naturally that some new
physics should govern the system and we may look some insight of the
yet unknown physics above $\Lambda$ under
the guide of compositeness conditions. 

The present model does not have supersymmetry in which 
fermions are automatically introduced.
We should take care of anomaly which is essential to cooperate with a
realization of hidden local symmetry \cite{moore}.
This issue is postponed for future work.

 Finally we would like to comment on the Berry phase \cite{Berry},
another approach to  dynamical gauge bosons. 
It is rather common phenomenon
that a gauge field is induced in a quantum mechanical system
which has redundant degrees of freedom.
 For example, in a molecule electrons work as hidden degrees of
freedom added to nuclei coordinates.
Electronic degrees are integrated out by the Born-Oppenheimer
approximation but leave an induced gauge field
which influences motion of nuclei.
Induced gauge fields are found also in quantum mechanics
of topologically nontrivial manifolds \cite{TT}.
However such kinds of gauge fields are not dynamical but static
configurations.
So it have been expected to find a mechanism to equip the induced
gauge fields with their own dynamical degrees of freedom.
Kikkawa \cite{kikkawa,kikkawaTamura}
provided such a mechanism using field theory with
extra compactified dimensions.
His argument is quite different from the present approach, but it
seems also interesting.
We hope that the dynamical generation of gauge bosons using the notion
of hidden local symmetry, together with the help of Berry phase
mechanism will 
provides some clue for dynamical origin 
of duality.

\section*{Acknowledgments}
We would like to thank Kugo for his valuable comments and
discussion.
Discussion with Harano, Imamura, Kikukawa, Maekawa, Sato, Sugimoto and
Takahashi is also helpful and encouraging.
The main part of this work was done during the Kashikojima Workshop 
held in August 1996 at Kashikojima Center,
which was supported by the Grand-in Aid for Scientific Research 
(No. 07304029) from the Ministry of Education, Science and Culture. 
M.B., Y.T. and S.T. are supported in part by the Grand-in Aid for
Scientific Research
(No. 33901-06640416) (No. 3113) (No. 08740200)
from the Ministry of Education, Science and Culture. 
Y.T. is a JSPS fellow.

\vspace{18mm}

%\newpage

%%%%%%%%%%%%%%%%%%%%%%%%%%%%%%%%%%%%%%%%%%%%%%%%%%%%%%%%%%%%%%%%%%%%%%
\newcommand{\J}[4]{{\it #1} {\bf #2} (19#3) #4}
\newcommand{\MPL}{Mod.~Phys.~Lett.}
\newcommand{\IJMP}{Int.~J.~Mod.~Phys.}
\newcommand{\NP}{Nucl.~Phys.}
\newcommand{\PL}{Phys.~Lett.}
\newcommand{\PR}{Phys.~Rev.}
\newcommand{\PRL}{Phys.~Rev.~Lett.}
\newcommand{\AP}{Ann.~Phys.}
\newcommand{\CMP}{Commun.~Math.~Phys.}
\newcommand{\PTP}{Prog. Theor. Phys.}
\newcommand{\Suppl}{Prog. Theor. Phys. Suppl.}
%%%%%%%%%%%%%%%%%%%%%%%%%%%%%%%%%%%%%%%%%%%%%%%%%%%%%%%%%%%%%%%%%%%%%%

\end{document}